\documentclass[reqno]{amsart} \usepackage{amscd}
\usepackage{epsfig}

\theoremstyle{definition}

\theoremstyle{remark} 
\numberwithin{equation}{section}
\newcommand{\be}{\begin{eqnarray}} 
\newcommand{\ee}{\end{eqnarray}}
\newcommand{\field}[1]{\ensuremath{\mathbb{#1}}}

\newcommand{\II}{\field{I}}

\newcommand{\RR}{\field{R}}

\DeclareMathOperator{\SO}{SO}
\DeclareMathOperator{\SU}{SU}

\newcommand{\lanln}[1]{$\langle$\texttt{arXiv:#1}$\rangle$}

\begin{document}

\title[Racah Coefficients]{$\SO_0(1,d+1)$ 
Racah coefficients: \\ Type I Representations}
\author{Kirill Krasnov} \address{School of Mathematical Sciences \\
  University of Nottingham \\ Nottingham, NG7 2RD, UK and
Bogolyubov Institute for Theoretical Physics \\ Metrologichna 14 b \\
  Kiev, 03143, Ukraine}
\author{Jorma Louko}\address{School of Mathematical Sciences \\
  University of Nottingham \\ Nottingham, NG7 2RD, UK}
\email{kirill.krasnov, jorma.louko at nottingham.ac.uk}
\begin{abstract} We use AdS/CFT inspired methods to study the 
Racah coefficients for type I representations of the Lorentz group
$\SO_0(1,d+1)$ with $d>1$. For such representations (a multiple of)
the Racah coefficient can be represented as an integral of a product
of 6 bulk-to-bulk propagators over 4 copies of the hyperbolic space
$H_{d+1}$. To compute the integrals we represent the bulk-to-bulk
propagators in terms of bulk-to-boundary ones. The bulk integrals can
be computed explicitly, and the boundary integrations are carried out
by introducing Feynman parameters. The final result is an integral
representation of the Racah coefficient given by 4 Barnes-Mellin type
integrals.
\end{abstract}
\maketitle
\section{Introduction and the Main Result}
\label{sec:intr}

Racah or Racah-Wigner (RW) coefficients or, as they are often called,
$6j$-symbols are important objects in group representation
theory. They depend on 6 irreducible representations of the group and
are obtained by multiplying 4 Clebsch-Gordan coefficients and summing
over the basis labels. Explicit formulas for RW coefficients are
available for the case of group $\SU(2)$. They are given by a (finite)
generalized hypergeometric series ${}_4 F_3(1)$ of unit argument
\cite{VK-1}. RW coefficients for other compact groups, especially the
unitary group ${\rm U}(n)$ and the rotation group $\SO(n)$ have also
been computed. For non-compact groups these coefficients have received
much less attention. This has to do with the envisaged physical
application of the RW coefficients: they figure prominently in the
lattice approach to QCD. As QCD concerns itself with compact gauge
groups such as $\SU(n)$ there is no field theoretic motivation to
compute the non-compact RW coefficients. This situation changed with
the introduction of the spin foam models of quantum gravity
\cite{Review}. Here of physical interest are exactly models with
non-compact groups such as the Lorentz group. These models use RW and
other similar coefficients in an essential way \cite{BC}. It thus
became of importance to study and derive explicit expressions for the
non-compact RW coefficients. Of particular interest are the Lorentz
group RW coefficients. This is the problem that is considered in the
present paper. Another obvious complication that arises in the
non-compact case is that the unitary representations are
infinite-dimensional. Sums of the compact case get replaced by
integrals; the problem requires a more careful analytical treatment,
including a careful analysis of the convergence of all the
expressions. This paper gives such a treatment for $\SO_0(1,d+1)$ RW
coefficients for type I representations, for $d>1$. Here
$\SO_0(1,d+1)$ denotes the connected component of $\SO(1,d+1)$. We
shall assume $d>1$ throughout the paper when not explicitly mentioned
otherwise.

Motivations for considering the RW coefficients for type I
representations are two-fold: (i) these are the representations that
are of importance for quantum gravity applications \cite{BC,FK}; (ii)
RW coefficients for type I representations can be obtained from a
certain integral over several copies of the hyperbolic space. As we
show in this paper, these integrals and thus the RW coefficients can
be given in terms of 4 Barnes-Mellin type integrals. One may be able
to use this integral representation for an analytic continuation, and
in this way obtain RW coefficients for other representations.

We recall here some basic facts about $\SO_0(1,d+1)$ representation
theory \cite{VK-2}. Representations of type I form the most degenerate
series of representations. At the same time they are the simplest and
the most studied ones. These irreducible unitary representations
appear in the decomposition of the space $L^2(H_{d+1})$ of square
integrable functions on the hyperbolic space $H_{d+1}$ into
irreducible representations.  Thus, type I representations can be
realized in the space of $L^2(H_{d+1})$ functions.  It is this
realization that will be the starting point for our analysis.

These representations can also be realized in the space of homogeneous functions on the light cone in 
Minkowski space $M^{1,d+1}$. We shall refer to minus degree of homogeneity by the conformal dimension
$\Delta$ of the representation. For representations of type I:
\be\label{type-1}
\Delta=d/2 + i \rho, 
\ \ 
\rho \in \mathbb{R}. 
\ee
There is another important series of unitary representations with real
integral $\Delta$. We shall not consider them here. We shall also
require a notion of the dual representation. Its conformal dimension
$\bar{\Delta}$ is such that:
\be\label{sum}
\Delta+\bar{\Delta}=d.
\ee
The dual representation is an equivalent representation.

A brief description of the logic of the paper is as follows. We
represent the RW coefficient as an integral over 4 copies of the
hyperbolic space of a product of 6 ``propagators'', the corresponding
expression to be given in the main text.  This way of representing the
$6j$-symbols was proposed in \cite{FK} and later explored in papers by
many authors, in particular \cite{BC} and \cite{Manuel}. In paper
\cite{BC} the bulk-to-bulk propagator was represented as a composition
of two ``bulk-to-boundary'' ones. This representation will be of
central importance for us in this paper. The terminology
``bulk-to-boundary'' and ``bulk-to-bulk'' is that used in the AdS/CFT
context \cite{Witten}.

Replacing each bulk-to-bulk propagator by a composition of two
bulk-to-boundary ones and using the usual field theoretic trick of
Feynman parameters, all the integrals over the hyperbolic space as
well as over the boundary can be taken. Moreover, as we shall see, all
the integrals over the Feynman parameters can be taken as well, at the
expense of introducing 4 Barnes-Mellin type integrals. The expression
we end up with has these 4 integrals remaining. Our final result for
the RW coefficient is:
\be\label{result}
(6\Delta)=
K_{\Delta_1,\Delta_2,\Delta_3,\Delta_5,\Delta_6} K_{\Delta_4,\bar{\Delta}_2,\bar{\Delta}_3,\bar{\Delta}_5,\bar{\Delta}_6}
\\ \nonumber
\int \frac{d\lambda ds}{(2\pi i)^2} \int \frac{d\lambda' ds'}{(2\pi i)^2} \Gamma(-s) \Gamma(-\lambda) \Gamma(-s') \Gamma(-\lambda') \\ \nonumber
 \Gamma(\gamma-s) \Gamma(\delta-\lambda) \Gamma(\alpha+s+\lambda)\Gamma(\beta+s+\lambda)\\ \nonumber
\Gamma(\gamma'-s') \Gamma(\delta'-\lambda') \Gamma(\alpha'+s'+\lambda')\Gamma(\beta'+s'+\lambda')\\ \nonumber
\pi^{d/2}  \frac{\Upsilon_d(\lambda-s'-\lambda'-A)\Upsilon_d(\lambda'-s-\lambda+A)}{\Upsilon_d(-s-s')}, 
\ee 
where
\be
\Upsilon_d(x)=\frac{\Gamma(x)}{\Gamma(d/2-x)}. 
\ee
The integration contour in each of the four variables is parallel to
the imaginary axis at $\mathrm{Re}(z) = r$, where the parameter $r$
can be chosen freely in the interval $-d/8<r<0$. 
Various quantities that appear in \eqref{result} are defined as follows:
\be\nonumber
&&
\alpha=\frac{\bar{\Delta}_1+\Delta_3-\Delta_2}{2}, \qquad \alpha'=\frac{\bar{\Delta}_4+\bar{\Delta}_6-\bar{\Delta}_2}{2},\\ 
\nonumber 
&&
\beta=\frac{\bar{\Delta}_1+\Delta_6-\Delta_5}{2}, 
\qquad \beta'=\frac{\bar{\Delta}_4+\bar{\Delta}_3-\bar{\Delta}_5}{2}, \\ 
\label{coef-res}
&&
\gamma=\frac{\Delta_2-\Delta_3+\Delta_5-\Delta_6}{2}, 
\qquad \gamma'=\frac{\bar{\Delta}_2-\bar{\Delta}_3+\bar{\Delta}_5-\bar{\Delta}_6}{2}, \\ 
\nonumber
&&
\delta=\frac{\Delta_1-\bar{\Delta}_1}{2}, \qquad \delta'=\frac{\Delta_4-\bar{\Delta}_4}{2}, \\ 
\nonumber
&&
A= \frac{\Delta_1-\Delta_4+\Delta_2+\Delta_5-d}{2},
\\ \nonumber 
&&
K_{\Delta_1,\Delta_2,\Delta_3,\Delta_5,\Delta_6}
\\ 
\nonumber
&& \ \ = 
\frac{\pi^{3d/2} \Gamma(\frac{\Delta_1+\Delta_2+\Delta_3-d}{2})\Gamma(\frac{\Delta_1+\Delta_5+\Delta_6-d}{2})
\Gamma(\frac{\Delta_2+\Delta_3-\Delta_1}{2}) \Gamma(\frac{\Delta_5+\Delta_6-\Delta_1}{2})}
{4 \Gamma^2(\Delta_1)\Gamma(\Delta_2)\Gamma(\Delta_3)\Gamma(\Delta_5)\Gamma(\Delta_6)}.
\ee
As is shown in the main text, for type I representations the
Barnes-Mellin integrals in \eqref{result} converge in absolute value,
so that this formula gives a well-defined representation of the RW
coeffcient.  The expression \eqref{result} can potentially be used for
numerical study of the $6j$-symbol as well as for the study of the
asymptotics. We do not consider these problems in the present paper.

Given the answer \eqref{result} for the RW coefficient one may ask if
it is ``as expected'' from general considerations. For instance, it
can be expected that the non-compact case answer can be obtained by
some ``analytic continuation'' from the one of the compact case. By
``analytic continuation'' one could mean some procedure of rather free
manipulation with the compact answer by replacing the finite sums with
integrals and continuing the representation labels to complex
values. For example, the answer for the $\SU(2)$ RW coefficient is
given by a single finite sum. Should one expect the $\SO_0(1,2)$, and
more generally $\SO_0(1,d+1)$, RW coefficient to be given by a single
integral? Here we would like to argue against this expectation.

As is clear from the computation
of the $6j$-symbol given in \cite{VK-1}, the fact that representations of $\SU(2)$ are highest
weight is used in an essential way. The idea of the computation is as follows. Consider a
product of 3 Clebsch-Gordan coefficients contracted to form a ``star-triangle''. When
the Clebsch-Gordan coefficients are appropriately normalized, this star-triangle is proportional
to a single Clebsch-Gordan coefficient with the proportionality coefficient being the $6j$-symbol:
\be
\sum_{m_{ij}} C^{l_1 l_{12} l_{13}}_{m_1 m_{12} m_{13}} C^{l_2 l_{23} l_{12}}_{m_2 m_{23} m_{12}} 
C^{l_3 l_{13} l_{23}}_{m_3 m_{13} m_{23}} = (6j) C^{l_1 l_2 l_3}_{m_1 m_2 m_3} .
\ee
Here the $m$-labels should satisfy a conservation law for each Clebsch-Gordan coefficient:
\be
m_1 = m_{12}+ m_{13},\, m_2= m_{23}+ m_{12},\, m_3 = m_{23} - m_{13},\, m_2 =m_1 + m_3.
\ee
Thus, there is only a single loop ``momentum'' to be summed over. Each Clebsch-Gordan coefficient
is also given by a finite sum, see \cite{VK-1}. Therefore, we have 4 sums on the left
hand side and one sum times the $6j$-symbol on the right hand side. What simplifies the computation
considerably is that one can set the basis labels $m_1, m_2$ to be highest weight. For
such values of $m$ the sum that gives the Clebsch-Gordan coefficient reduces to a single term.
Thus, if we put two of the external states to be highest weight, we get only two sums
remaining on the left hand side and no sum times the $6j$ on the right. One of the summations
on the left hand side can be carried out explicitly using Gauss's sum. One gets
an expression for the $6j$ given by a single sum. 

Let us now return to the non-compact case. Going to a ``momentum''
basis, analogous to the $m$-basis of $\SU(2)$, one can have momentum
conserved at each vertex.  One analogously will have to do only a
single integral over the loop momentum. However, in the momentum basis
the Clebsch-Gordan coefficients are no longer as simple as in
\eqref{3j}. One should expect an integral representation with at least
one integral. Now, the type I representations are not highest or
lowest weight. Thus, there is no simplification of the momentum basis
Clebsch-Gordan coefficients that can be used in the computation of the
$6j$-symbol. The total number of integrals remains at least 4 on the
right hand side and at least one on the left hand side.  Thus, at the
very best one expects the non-compact $6j$ symbol for type I
representations to be given by 3 integrals. We have tried to further
simplify our answer \eqref{result} in line with this this
expectation. Unfortunately, no summation theorems for the generalized
hypergeometric series seem to be applicable, and we were not able to
simplify the result further. It thus remains an interesting open
problem to find a simpler representation for the RW coefficient than
the one given by \eqref{result}.

This paper is organized as follows. We start in section \ref{sec:prop}
by reviewing the bulk-to-bulk and bulk-to-boundary
propagators. Section \ref{sec:rw} gives a formula for the RW
coefficients in terms of bulk-to-bulk propagators, as well as another
convenient representation in terms of a boundary 4-point function.  In
section \ref{sec:3j} we compute the 3-point function, and in section
\ref{sec:norm} we discuss its normalization. An important regularization
procedure is introduced here. In section
\ref{sec:4-point} we compute the 4-point function. 
Finally, in section \ref{sec:comp} we compute the
remaining sphere integrals. 
Certain technical results are given in two appendices.

\section{Bulk-to-bulk and bulk-to-boundary propagators}
\label{sec:prop}

In this section we allow $d$ to be any positive integer. 

We consider the $d$-dimensional conformal group $\SO_0(1,d+1)$. 
One of the associated homogeneous spaces is the hyperbolic space
$H_{d+1}=\SO_0(1,d+1)/\SO(d+1)$. The space $L^2(H_{d+1})$ of square integrable functions on $H_{d+1}$ decomposes into 
irreducible representations of the so-called type I series. 
More generally, the most degenerate series of representations of the conformal group is that
in the space of homogeneous functions on the light cone in Minkowski space $M^{1,d+1}$. Functions of degree
of homogeneity $-\Delta$ form an irreducible representation of ``conformal dimension'' $\Delta$. Not all
of these representations are unitary. For unitary representations of type I the conformal dimension is given by
\eqref{type-1}.

Thoughout the paper, we label the representations by their conformal dimension~$\Delta$. 
While the final results will be justified only for the type I values \eqref{type-1}, 
several intermediate results remain valid for more general values, and we present the 
intermediate results in this form in view of potential extensions beyond type I representations. 

To proceed, we need the notions of bulk-to-bulk and bulk-to-boundary
propagators \cite{Witten}. 
We use the
upper half-space model of $H_{d+1}$. Let the coordinates 
in the upper half-space be $(\xi_0>0,\xi_i), i=1,\ldots,d$. 
The metric is then:
\be
ds^2 = \frac{1}{\xi_0^2}(d\xi_0^2 + \sum_i d\xi_i^2).
\ee
The boundary of $H_{d+1}$ is the set of points with $\xi_0=0$ and the point at infinity. 
When referring to points of the boundary we shall use
a different letter $x: x_i=\xi_i$.

Following \cite{Witten}, we introduce the following function of a
point on the boundary 
$x$ and a bulk point $\xi$ in $H_{d+1}$:
\be\label{bound-prop}
K_\Delta(\xi,x) = \frac{\xi_0^\Delta}{(\xi_0^2+|\xi-x|^2)^\Delta}.
\ee
We refer to this function as the bulk-to-boundary propagator. 
Another, more familiar from mathematics literature, expression for this object is given by:
\be\label{bound-prop-1}
K_\Delta(\hat\xi,\hat x)=(\hat \xi \cdot \hat x)^{-\Delta},
\ee
where the vectors $\hat\xi$ and $ \hat x$ in  Minkowski space $M^{1,d+1}$ are the representatives of 
respectively $\xi$ and $x$ in the hyperboloid model of $H_{d+1}$
and the dot product is the Minkowski pairing.  
$\hat\xi$ is unit timelike and $ \hat x$ is null.  

The bulk-to-bulk propagator is obtained by taking a product of two propagators 
\eqref{bound-prop}, one for representation $\Delta$, another for the dual representation $\bar{\Delta}$, 
and integrating over the point on the boundary: 
\be\label{bulk-prop}
K_\Delta(\xi_1,\xi_2)= \int_{S^d} d^dx\, K_\Delta(\xi_1,x) K_{\bar{\Delta}}(\xi_2,x). 
\ee
We have denoted the integration domain, $x\in\mathbb{R}^d$, by $S^d$ as a reminder 
of the boundary topology in the Poincare ball model of $H_{d+1}$. 
As the integrand is asymptotic to a constant times $|x|^{-2d}$ at $|x|\to\infty$, 
the integral converges for all $\Delta\in\mathbb{C}$. 

The bulk-to-bulk propagator \eqref{bulk-prop} can be computed explicitly. 
To begin, assume 
${\rm Re}(\Delta) >0$ and 
${\rm Re}(\bar\Delta) >0$. Using the Feynman representation reviewed in Appendix~\ref{app:integrals}, 
we obtain: 
\be
K_\Delta(\xi_1,\xi_2)= \frac{(\xi_1^0)^\Delta (\xi_2^0)^{\bar{\Delta}}}{\Gamma(\Delta)\Gamma(\bar{\Delta})} 
\int_{S^d} d^dx \int dt du\,\, t^{\Delta-1} u^{\bar{\Delta}-1} \\ \nonumber
 e^{-t (\xi_1^0)^2 -u (\xi_2^0)^2 - t|\xi_1-x|^2 -u|\xi_2-x|^2}.
\ee
Taking the integral over $x$ yields: 
\be\nonumber
K_\Delta(\xi_1,\xi_2)= \frac{(\xi_1^0)^\Delta (\xi_2^0)^{\bar{\Delta}}}{\Gamma(\Delta)\Gamma(\bar{\Delta})} 
\int dt du\,\, t^{\Delta-1} u^{\bar{\Delta}-1} \\ \nonumber
e^{-t (\xi_1^0)^2 -u (\xi_2^0)^2} \frac{\pi^{d/2}}{(t+u)^{d/2}} e^{-\frac{t u}{t+u} |\xi_1-\xi_2|^2}.
\ee
By construction, $K_\Delta(\xi_1,\xi_2)$ is invariant under the action of the 
Lorentz group and hence depends on $\xi_1$ and $\xi_2$ only through their hyperbolic distance~$l$. 
Using this invariance, we can choose $\xi_1 = (\xi_1^0, 0)$ and 
$\xi_2 = (\xi_2^0, 0)$, in which case 
\be
l=\log{(\xi_1^0/\xi_2^0)}.
\ee
Writing $\mu=e^l$ and rescaling the integration variables, we obtain: 
\be
K_\Delta(\xi_1,\xi_2)= \frac{\pi^{d/2} \mu^\Delta}{\Gamma(\Delta)\Gamma(\bar{\Delta})}
\int \frac{dt du}{(t+u)^{d/2}} \,\, t^{\Delta-1} u^{\bar{\Delta}-1} e^{-t \mu^2-u}.
\ee
The integrals over $t$ and $u$ can now be taken with the change of variables 
\be\label{change}
\lambda = \frac{t}{t+u}, \qquad dt \, du = \frac{u d\lambda \, du}{(1-\lambda)^2}, 
\ee
where $0<\lambda<1$. We get:
\be
K_\Delta(\xi_1,\xi_2)= \frac{\pi^{d/2} \mu^\Delta}{\Gamma(\Delta)\Gamma(\bar{\Delta})}
\int_0^1 \frac{d\lambda}{(1-\lambda)^{2-d/2}} \int_0^\infty \frac{du}{u^{d/2-1}}\\ \nonumber
\left(\frac{\lambda u}{1-\lambda}\right)^{\Delta-1} u^{\bar{\Delta}-1} 
e^{-u(\frac{\lambda \mu^2}{1-\lambda}+1)}.
\ee 
The integral over $u$ gives: 
\be
K_\Delta(\xi_1,\xi_2)= \frac{\pi^{d/2} \mu^\Delta \Gamma(d/2)}{\Gamma(\Delta)\Gamma(\bar{\Delta})}
\int_0^1 d\lambda\, \lambda^{\Delta-1} (1-\lambda)^{d-\Delta-1} (1-\lambda(1-\mu^2))^{-d/2}.
\ee
Using the integral representation of the hypergeometric function, 
\be
F(a,b,c;z)=\frac{\Gamma(c)}{\Gamma(b)\Gamma(c-b)} \int_0^1 dt\,\, t^{b-1} (1-t)^{c-b-1} (1-tz)^{-a},
\ee
we finally obtain: 
\be
\label{eq:bulk-bulk-hyper} 
K_\Delta(\xi_1,\xi_2)= \frac{\pi^{d/2} \mu^\Delta \Gamma(d/2)}{\Gamma(d)} 
F(d/2,\Delta,d,1-\mu^2).
\ee
By analytic continuation, the result \eqref{eq:bulk-bulk-hyper} 
holds for all $\Delta\in\mathbb{C}$. 
Note that there is no singularity when $\Delta$ is a non-positive integer; 
the hypergeometric series just terminates then. We also note that,
in view of the identity:
\be
F(c-a,c-b,c;z)=(1-z)^{a+b-c}F(a,b,c;z)
\ee
we have $K_\Delta(\xi_1,\xi_2)=K_{\bar{\Delta}}(\xi_1,\xi_2)$. Thus,
the bulk-to-bulk propagator is a non-oriented one. Indeed, the change of
orientation is equivalent to the replacement $\Delta\to\bar{\Delta}$, but
this does not change the propagator. 

The result \eqref{eq:bulk-bulk-hyper} can be expressed in terms of a 
Legendre function~\cite{abra-stegun}. For $d=1$, the formula was 
given in \cite{Manuel}. For $d=2$, the result can be given in terms of an elementary function as 
\be
K_\Delta(\xi_1,\xi_2)= \frac{\pi}{\Delta-1}
\frac{\sinh{(\Delta-1)l}}{\sinh{l}}.
\ee
For type I representations this reduces to: 
\be
K_\rho(\xi_1,\xi_2)= \frac{\pi}{\rho}
\frac{\sin{\rho l}}{\sinh{l}}.
\ee

\section{Racah-Wigner coefficients}
\label{sec:rw}

We shall compute the RW coefficients 
through an object called the 
$(6\Delta)$ symbol, given by 
an integral over 4 copies of $H_{d+1}$ of a product of 6 bulk-to-bulk propagators:
\be\label{6j}
(6\Delta)=\int_{H_{d+1}} d\xi_1 \ldots d\xi_4 \prod_{i<j} K_{\Delta_{ij}}(\xi_i,\xi_j).
\ee
Here $i,j=1,\ldots,4$ enumerate the points integrated over and
$\Delta_{ij}$ are the 6 representations that the $(6\Delta)$ symbols
depend upon. We begin formally with general $\Delta_{ij}\in\mathbb{C}$
but will eventually specialise to the type I values
\eqref{type-1}. Since the non-type-I representations are not
realizable in $L^2(H_{d+1})$, we do not expect that the integral
\eqref{6j} can be made convergent for them even after eliminating the
infinite volume factor by the procedure of section~\ref{sec:norm}.

The relation between the $(6\Delta)$ symbol 
(\ref{6j}) and the RW coefficients has been discussed in~\cite{FK}. 
To make the present paper
self-contained, we shall briefly recall the relation here. 

By definition, the representations of type I 
(with respect to a subgroup $H$) are 
those which contain a vector invariant under $H$. When 
$G=\SO_0(1,d+1)$, the subgroup $H$ is taken to be the maximal
compact subgroup $H=\SO(d+1)$. Let $\pi^\rho$ be a type I 
representation, labelled by the index $\rho$, and let $\omega$ be an
$H$-invariant vector. We define on $G$ the function $\phi^\rho$ as the
$\omega\omega$ matrix element of $\pi^\rho$, 
\be
\label{eq:phisuprho-def}
\phi^\rho(g) := t^\rho_{\omega\omega}(g) :=  \langle \omega |
\pi^\rho(g) |\omega \rangle.
\ee
$\phi^\rho$ is called a {\it spherical} function. 
Given $\phi^\rho$, we define the kernel 
\be
\label{eq:Krho-spherdef}
K_\rho(g_1,g_2) := \phi^\rho(g_1^{-1} g_2).
\ee
As $\omega$ is invariant under $H$, (\ref{eq:phisuprho-def}) shows
that $\phi^\rho$ projects into a function on the double coset
$H\backslash G / H$, and (\ref{eq:Krho-spherdef}) then shows that the
kernel projects into a function on two copies of the homogeneous space
$G/H$. In the case of $G=\SO_0(1,d+1)$ and $H=\SO(d+1)$, this
homogeneous space is the hyperbolic space $H^{d+1}$, and from the explicit evaluation of the spherical function in section 9.3.1 of \cite{VK-2}
it is seen that the kernel 
(\ref{eq:Krho-spherdef}) concides (up to a multiple that depends
on the normalization of~$\omega$) with the 
bulk-to-bulk propagator~(\ref{eq:bulk-bulk-hyper}). 

Next, we recall the definition of the Clebsch-Gordan coefficients. 
Given two (irreducible) representations $V^{\rho_1},
V^{\rho_2}$ of $G$, one can decompose the tensor product representation
$V^{\rho_1} \otimes V^{\rho_2}$ into irreducible
representations $V^\rho$ as 
\be
\label{eq:Vrho1-tensor-Vrho2}
V^{\rho_1} \otimes V^{\rho_2} = \oplus_{\rho_3} n(\rho_3) V^{\rho_3},
\ee
where $n(\rho)$ are multiplicities. 
The Clebsch-Gordan coefficients are the matrix elements of the
intertwining map defined by this decomposition. Explicitly, if 
$i_1$ and $i_2$ are vectors in respectively $V^{\rho_1}$ and
$V^{\rho_2}$, we have 
\be
\label{eq:CG-def}
|i_1 \rangle \otimes \,\, |i_2\rangle 
= 
\sum_{\alpha \rho_3 i_3} {}^\alpha C^{\rho_3 i_3}_{\rho_1 i_1 \rho_2
i_2}\,\, |i_3 \rangle,
\ee
where $\{i_3\}$ is a (generalised) basis in $V^{\rho_3}$, 
the sum on the right-hand side includes a sum over the multiplicity
label $\alpha = 1\ldots n(\rho_3)$, and the sum over the other labels
is understood in the sense of a sum or an integral depending on
whether the labels are discrete or continuous. 
The $C$'s on the right-hand side are then the Clebsch-Gordan
coefficients in the respective bases. 

We will further need the orthogonality relation 
for the matrix elements 
$t^\rho_{i_1 i_2}(g) :=  \langle i_1 |
\pi^\rho(g) | i_2 \rangle$. 
This relation reads 
\be
\label{eq:me-orthog}
\int dg \,\, t^{\rho_1}_{i_1 j_1}(g) t^{\rho_2}_{i_2 j_2}(g) =
\delta^{\rho_1 \rho_2} \delta_{i_1 i_2} \delta_{j_1 j_2}, 
\ee 
where we have used the reality of the matrix elements
for $G=\SO_0(1,d+1)$, and the 
deltas on the right-hand side are understood as Kronecker
or Dirac deltas as appropriate. 
For a compact group, whose representations are finite-dimensional, 
the right-hand side of 
(\ref{eq:me-orthog}) would usually be normalised to include the factor
$1/{\rm dim}(j_1)$, but in our
case of a noncompact group the representations are
infinite-dimensional and such a normalisation is not an option. 
Using 
(\ref{eq:CG-def}) and 
(\ref{eq:me-orthog}), 
it is then possible to compute the integral of the
product of three matrix elements, with the result 
\be
\label{eq:3-prod-int}
\int dg\,\, t^{\rho_1}_{i_1 j_1}(g) t^{\rho_2}_{i_2 j_2}(g) t^{\rho_3}_{i_3 j_3}(g) =
\sum_{\alpha} {}^\alpha C^{\rho_3 i_3}_{\rho_1 i_1 \rho_2 i_2} {}^\alpha C^{\rho_3 j_3}_{\rho_1 j_1 \rho_2 j_2}.
\ee

Now, when the the multiplicity of type I representations in the tensor
product 
(\ref{eq:Vrho1-tensor-Vrho2})
of two such representations is equal to one, 
the multiplicity label on the type I Clebsch-Gordan coefficients becomes
redundant, and it can be shown from 
(\ref{eq:3-prod-int}) that 
the $(6\Delta)$ symbol 
\eqref{6j} equals the usual type I 
RW coefficient, constructed from the Clebsch-Gordan
coefficients, multiplied by four factors of the form 
$C^{\rho_3 \omega}_{\rho_1 \omega \rho_2 \omega}$, one
such factor coming from each $\xi$ in~\eqref{6j}.
In this case one can therefore regard the 
$(6\Delta)$ symbol 
\eqref{6j} as the type I RW coefficient up to a certain normalisation
factor. 
When the multiplicity of type I representations in the tensor
product
of two such representations is not equal to one, 
on the other hand, 
the multiplicity label
on the type I Clebsch-Gordan coefficients cannot be dropped. The 
$(6\Delta)$ symbol 
\eqref{6j} involves then factors of the form 
${}^\alpha C^{\rho_3 \omega}_{\rho_1 \omega \rho_2 \omega}$ 
for each~$\xi$, and there is a \emph{sum\/} over
the four independent multiplicity labels. 
As the type I RW coefficients then have a multiplicity label corresponding
to each~$\xi$, the $(6\Delta)$ symbol 
\eqref{6j} is no longer a simple multiple of the type I RW coefficients. 

Thus, to establish the relation of the $(6\Delta)$ symbol 
\eqref{6j} to the usual
RW coefficients it remains to find the multiplicity of type I 
representations in the tensor product
(\ref{eq:Vrho1-tensor-Vrho2}) of two such representations. For
$\SO_0(1,2)$ this multiplicity is known to be 2 \cite{Puk}. The
$\SO_0(1,2)$ $(6\Delta)$ symbol \eqref{6j} is therefore not a simple
multiple of (any of the) RW coefficients. The $\SO_0(1,2)$ $(6\Delta)$
symbol \eqref{6j} can still be computed and used as a basis of a state
sum model~\cite{Manuel}, 
but its group theoretic interpretation does not appear straightforward. 

For
$\SO_0(1,d+1)$ with $d>1$, 
the multiplicity of type I 
representations in the tensor product
(\ref{eq:Vrho1-tensor-Vrho2}) of two such representations is one. 
To see this, we recall that the principal series representations can be obtained as the induced representations of the parabolic subgroup 
$MAN$, where $N$ is the $d$-dimensional Abelian group of null rotations, $A$ is the one-dimensional group of dilatations and 
$M=\SO(d)$, and in particular the type I representations are induced from the \emph{trivial\/} representation of $M$ \cite{Dobrev}. 
The claim then follows from Theorem 
16 of \cite{Martin} and the discussion of $\SO_0(1,d+1)$ 
on pages 182--183 therein. This establishes the 
relation of our $(6\Delta)$ symbol \eqref{6j} 
to the usual RW coefficients for 
$\SO_0(1,d+1)$ with $d>1$. 

\begin{figure}
\centering
\epsfig{figure=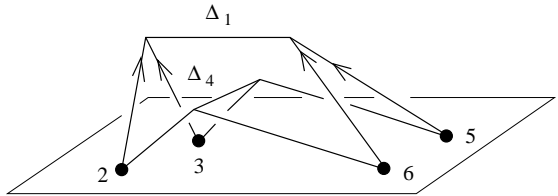, height=1in}
\caption{Expression for the Racah-Wigner coeffient in terms of the 4-point functions.\label{fig-2}}
\end{figure}

To proceed with the analysis of the $(6\Delta)$ symbol, we represent four of the six bulk-to-bulk propagators in \eqref{6j}
as in \eqref{bulk-prop}, while leaving the two bulk-to-bulk propagators corresponding to the opposite edges of 
the tetrahedron in their original form. By doing this one achieves the representation as is shown in Fig.~\ref{fig-2}.

Thus, let us introduce the following object:
\be\label{4-point}
D_{\Delta_1,\Delta_2,\Delta_3,\Delta_5,\Delta_6}(x_2,x_3,x_5,x_6) = \\ \nonumber \int_{H_{d+1}} d\xi d\eta \,\, 
K_{\Delta_2}(\xi,x_2)K_{\Delta_3}(\xi,x_3) K_{\Delta_1}(\xi,\eta) K_{\Delta_5}(\eta,x_5)K_{\Delta_6}(\eta,x_6).
\ee
This quantity is a particular 4-point function on the boundary. With the
help of this 4-point function, the RW coefficient can be written as:
\be\label{6j-4-point}
(6\Delta)=\int_{S^d} d^dx_2 d^dx_3 d^dx_5 d^dx_6 \,\, \\ \nonumber D_{\Delta_1,\Delta_2,\Delta_3,\Delta_5,\Delta_6}(x_2,x_3,x_5,x_6) 
D_{\Delta_4,\bar{\Delta}_2,\bar{\Delta_6},\bar{\Delta}_5,\bar{\Delta}_3}(x_2,x_6,x_5,x_3).
\ee
This representation is much more convenient than the one given by
\eqref{6j}, since the 4-point function
$D_{\Delta_1,\Delta_2,\Delta_3,\Delta_5,\Delta_6}(x_2,x_3,x_5,x_6)$
can be computed explicitly.  Computing the RW coefficient then reduces
to the computation of the $S^d$ integrals in
\eqref{6j-4-point}. As we shall 
see in section~\ref{sec:norm}, 
after a certain gauge fixing that is necessary to render the result finite,
what one has is actually not 4 integrals but a single integral over the sphere. 
This makes the representation \eqref{6j-4-point}
very convenient for practical applications. We note that this representation is only available in the
context of non-compact conformal groups, when there is a boundary and the associated representation
\eqref{bulk-prop}. For compact groups like $\SO(n)$ one can also compute the RW coefficients
for type I representations by integrating over the appropriate homogeneous manifold. However, 
there is no analog of the representation \eqref{6j-4-point}.

Thus, we would like to compute the 4-point function given by \eqref{4-point}. In order to do this, we use the representation
\eqref{bulk-prop} for the remaining bulk propagator. Thus, we represent the 4-point function as:
\be
D_{\Delta_1,\Delta_2,\Delta_3,\Delta_5,\Delta_6}(x_2,x_3,x_5,x_6) = \\ \nonumber \int d^dx_1\, C_{\Delta_1,\Delta_2,\Delta_3}(x_1,x_2,x_3)
C_{\bar{\Delta}_1,\Delta_5,\Delta_6}(x_1,x_5,x_6),
\ee
where we have defined the 3-point function:
\be\label{3-point}
C_{\Delta_1,\Delta_2,\Delta_3}(x_1,x_2,x_3) = \int_{H_{d+1}} d\xi K_{\Delta_1}(\xi,x_1) K_{\Delta_2}(\xi,x_2) K_{\Delta_3}(\xi,x_3).
\ee
We need to find this 3-point function.

\section{3-point function}
\label{sec:3j}

We note in passing that the 3-point function 
\eqref{3-point}
is the 
Clebsch-Gordan coefficient 
when the representations are realised in the space of functions on the boundary. 
Since $d>1$ by assumption, 
these Clebsch-Gordan coefficients have no multiplicity indices. 

The conditions for the integral in \eqref{3-point} to converge are 
\be\label{eq:conv1} 
{\rm Re}(\sum_i \Delta_i) > d 
\ee
and 
\be\label{eq:conv2} 
{\rm Re}(\Delta_1+\Delta_2-\Delta_3)>0,{\rm Re}(\Delta_2+\Delta_3-\Delta_1)>0,{\rm Re}(\Delta_1+\Delta_3-\Delta_2)>0.
\ee
These conditions can be conveniently obtained by transforming 
to the Poincare ball model of $H_{d+1}$: 
\eqref{eq:conv2} comes from neighbourhoods of the three boundary points $x_i$, and 
\eqref{eq:conv1} comes from the remaining part of the boundary. Note that \eqref{eq:conv2}
implies ${\rm Re}(\Delta_i)>0$ for all~$i$. Note also that the convergence holds for 
type I representations. 

In order to compute the bulk integral we use the Feynman parametrization 
\eqref{feyn} for each of the bulk-to-boundary propagators.
We get:
\be\label{3j-1}
C_{\Delta_1,\Delta_2,\Delta_3}(x_1,x_2,x_3) = \frac{1}{\Gamma(\Delta_1)}  \frac{1}{\Gamma(\Delta_2)} \frac{1}{\Gamma(\Delta_3)}
\int_0^\infty dt_1 dt_2 dt_3\, t_1^{\Delta_1-1} t_2^{\Delta_2-1} t_3^{\Delta_3-1} \\ \nonumber
\int_0^\infty \frac{d\xi_0}{\xi_0^{d+1}}\, \xi_0^{\sum_i \Delta_i} \int_{S^d} d^d\xi \,
e^{-t_1(\xi_0^2+|\xi-x_1|^2)-t_2(\xi_0^2+|\xi-x_2|^2)-t_3(\xi_0^2+|\xi-x_3|^2)}.
\ee
We now use the formulas \eqref{1}, \eqref{2} of Appendix \ref{app:integrals} 
to get:
\be
C_{\Delta_1,\Delta_2,\Delta_3}(x_1,x_2,x_3) = \frac{\pi^{d/2} \Gamma(\frac{\sum_i \Delta_i-d}{2})}{2 \Gamma(\Delta_1)\Gamma(\Delta_2)\Gamma(\Delta_3)}
\int_0^\infty dt_1 dt_2 dt_3\, t_1^{\Delta_1-1} t_2^{\Delta_2-1} t_3^{\Delta_3-1} \\ \nonumber
(S_t)^{-(\sum_i \Delta_i)/2} e^{-\frac{1}{S_t}(\sum_{i<j} t_i t_j |x_i-x_j|^2)}.
\ee

We now make a series of changes of variables of integration. The first change is:
\be\label{change-1}
t_i = (S_t)^{1/2} t_i' = (\sum_j t_j') t_i', \qquad {\rm det}\left( \frac{\partial t_i}{\partial t_j'}\right) = 2 (S_t)^{3/2}.
\ee
Removing the primes, we get:
\be
&{}& C_{\Delta_1,\Delta_2,\Delta_3}(x_1,x_2,x_3) \\ \nonumber &=& 
\frac{\pi^{d/2} \Gamma(\frac{\sum_i \Delta_i-d}{2})}{\Gamma(\Delta_1)\Gamma(\Delta_2)\Gamma(\Delta_3)}
\int_0^\infty dt_1 dt_2 dt_3\, t_1^{\Delta_1-1} t_2^{\Delta_2-1} t_3^{\Delta_3-1} e^{- \sum_{i<j} t_i t_j x_{ij}^2 }.
\ee
Here we have introduced:
\be
x_{ij}=|x_i-x_j|.
\ee

The second change of variables is:
\be\label{change-2}
t_1 t_2 \to \frac{t_1 t_2}{x_{12}^2}, \qquad t_1 t_3 \to \frac{t_1 t_3}{x_{13}^2}, \qquad t_2 t_3 \to \frac{t_2 t_3}{x_{23}^2}.
\ee
The integral reduces to:
\be
C_{\Delta_1,\Delta_2,\Delta_3}(x_1,x_2,x_3)=
\frac{1}
{(x_{12})^{\Delta_1+\Delta_2-\Delta_3} 
(x_{13})^{\Delta_1+\Delta_3-\Delta_2} (x_{23})^{\Delta_2+\Delta_3-\Delta_1}} \\ \nonumber
\frac{\pi^{d/2} \Gamma(\frac{\sum_i \Delta_i-d}{2})}{\Gamma(\Delta_1)\Gamma(\Delta_2)\Gamma(\Delta_3)}
\int_0^\infty dt_1 dt_2 dt_3\, t_1^{\Delta_1-1} t_2^{\Delta_2-1} t_3^{\Delta_3-1} e^{- \sum_{i<j} t_i t_j  }.
\ee

It is now possible to take the remaining integral in Feynman parameters by the following change of variables:
\be\label{change-3}
t_1 t_2 =u_3, \qquad t_1 t_3=u_2, \qquad t_2 t_3 = u_1,
\ee
so that:
\be
t_1^2 = \frac{u_2 u_3}{u_1}, \qquad t_2^2 = \frac{u_1 u_3}{u_2}, \qquad t_3^2 = \frac{u_1 u_2}{u_3}, \qquad 
{\rm det}\left( \frac{\partial t_i}{\partial u_j}\right) = \frac{1}{2\sqrt{u_1 u_2 u_3}}.
\ee
The integral over $t_i$ thus reduces to:
\be
\frac{1}{2}\int_0^\infty du_1 du_2 du_3 \, u_1^{\frac{\Delta_2+\Delta_3-\Delta_1-2}{2}} u_2^{\frac{\Delta_1+\Delta_3-\Delta_2-2}{2}} 
u_3^{\frac{\Delta_1+\Delta_2-\Delta_3-2}{2}} e^{-u_1-u_2-u_3} = \\ \nonumber
\frac{1}{2} \Gamma(\frac{\Delta_2+\Delta_3-\Delta_1}{2}) \Gamma(\frac{\Delta_1+\Delta_3-\Delta_2}{2}) 
\Gamma(\frac{\Delta_1+\Delta_2-\Delta_3}{2}).
\ee
Thus, we get for the Clebsch-Gordan coefficients:
\be\label{3j}
C_{\Delta_1,\Delta_2,\Delta_3}(x_1,x_2,x_3)=\frac{C(\Delta_1,\Delta_2,\Delta_3)}{(x_{12})^{\Delta_1+\Delta_2-\Delta_3} 
(x_{13})^{\Delta_1+\Delta_3-\Delta_2} (x_{23})^{\Delta_2+\Delta_3-\Delta_1}},
\ee
where 
\be\label{C}
&{}& C(\Delta_1,\Delta_2,\Delta_3) = \\ \nonumber &{}& \frac{\pi^{d/2} \Gamma(\frac{\Delta_1+\Delta_2+\Delta_3-d}{2})
\Gamma(\frac{\Delta_2+\Delta_3-\Delta_1}{2}) \Gamma(\frac{\Delta_1+\Delta_3-\Delta_2}{2}) 
\Gamma(\frac{\Delta_1+\Delta_2-\Delta_3}{2})}{2 \Gamma(\Delta_1)\Gamma(\Delta_2)\Gamma(\Delta_3)}.
\ee

\section{Normalization of the Clebsch-Gordan coefficients. 
Gauge fixing procedure.}
\label{sec:norm}

In this section we introduce a certain gauge fixing procedure that is
necessary to render the integrals defining the RW coefficient
finite. We do this considerering as an example the question of
normalization fo the Clebsch-Gordan coefficient. To this end, let us
compute the so-called theta symbol given by:
\be\label{theta}
\theta(\Delta_1,\Delta_2,\Delta_3) = \int_{H_{d+1}} d\xi_1 d\xi_2 \, 
K_{\Delta_1}(\xi_1,\xi_2) K_{\Delta_2}(\xi_1,\xi_2)K_{\Delta_3}(\xi_1,\xi_2).
\ee
Using the expression for the 3$j$ symbol 
that we have just computed, this becomes:
\be\label{theta-1}
\theta(\Delta_1,\Delta_2,\Delta_3) = C(\Delta_1,\Delta_2,\Delta_3) 
C(\bar{\Delta}_1,\bar{\Delta}_2,\bar{\Delta}_3)
\int_{S^d} d^dx_1 d^dx_2 d^dx_3 \frac{1}{x_{12}^d x_{13}^d x_{23}^d}.
\ee
The integral over $x_1,x_2,x_3$ in \eqref{theta-1} is divergent. This
divergence is a manifestation of the obvious divergence in integrating
over $H_{d+1}$ a function that is invariant under the isometries: The
integral is proportional to the infinite volume of $H_{d+1}$. As the
volume of $H_{d+1}$ is related to the volume of the Lorentz group, the
divergence ultimately comes from the fact that the Lorentz group is
non-compact.

In \eqref{theta}, the
divergence can be removed by fixing one of the integration points and
only integrating over the remaining one. To remove the divergence in
\eqref{theta-1}, and in similar integrals with more integration
points, we can use the action of the conformal group on the sphere to
put three of the integration points to specific locations. 
For $d=2$ this procedure completely fixes the $\SO_0(1,d+1)$ invariance
and hence eliminates the divergence \cite{Strings}. For $d>2$ this
procedure fixes the $\SO_0(1,d+1)$ invariance up to the $\SO(d-1)$
subgroup that leaves the three prescribed points invariant, but as
this subgroup is compact, the divergence has again been eliminated.

To implement this procedure, one needs to find the Faddeev-Popov
determinant that makes the result independent of the locations to
which the three integration points are fixed. 
We shall do this in Appendix \ref{app:measure}
by 
decomposing the Haar measure
on $\SO_0(1,d+1)$ in terms of the locations of the three fixed points
and the invariantly-normalised volume of the compact subgroup
$\SO(d-1)$ (which is trivial for $d=2$). 
The result is the replacement 
\be\label{measure-d}
\prod_i d^dx_i \to |x_A-x_B|^d |x_A-x_C|^d |x_B-x_C|^d \\ \nonumber
\delta^d(x_A-x_A^0)\delta^d(x_B-x_B^0)\delta^d(x_C-x_C^0) \prod_i
d^dx_i, 
\ee
where $x_A$, $x_B$ and $x_C$ are the three integration points that are
fixed respectively to the locations $x_A^0$, $x_B^0$ and~$x_C^0$. 
For $d=2$, formula 
\eqref{measure-d} reduces to that found in~\cite{Strings}. 

In the divergent integral~\eqref{theta-1}, 
the gauge fixing \eqref{measure-d} 
simply removes the integral. This gives for the 
normalization of the Clebsch-Gordan coefficients the formula 
\be
\label{theta-norm}
\theta(\Delta_1,\Delta_2,\Delta_3) = 
C(\Delta_1,\Delta_2,\Delta_3) 
C(\bar{\Delta}_1,\bar{\Delta}_2,\bar{\Delta}_3).
\ee

\section{Computation of the 4-point function}
\label{sec:4-point}

Using the expression \eqref{3j} for the 3-point function, the 4-point function becomes:
\be\label{eq:DCC-int}
D_{\Delta_1,\Delta_2,\Delta_3,\Delta_5,\Delta_6}(x_2,x_3,x_5,x_6) =  C_{\Delta_1,\Delta_2,\Delta_3}
C_{\bar{\Delta}_1,\Delta_5,\Delta_6} \int_{S^d} d^dx_1 \\ \nonumber \frac{1}{(x_{12})^{\Delta_1+\Delta_2-\Delta_3} 
(x_{13})^{\Delta_1+\Delta_3-\Delta_2} (x_{23})^{\Delta_2+\Delta_3-\Delta_1}}  \\ \nonumber \frac{1}{(x_{15})^{\bar{\Delta}_1+\Delta_5-\Delta_6} 
(x_{16})^{\bar{\Delta}_1+\Delta_6-\Delta_5} (x_{56})^{\Delta_5+\Delta_6-\bar{\Delta}_1}}.
\ee
In addition to the convergence conditions of section \ref{sec:3j} for each of the 3-point functions, the conditions for the integral in \eqref{eq:DCC-int} to converge are: 
\be
{\rm Re}(\Delta_1+\Delta_2-\Delta_3)<d, \qquad {\rm Re}(\Delta_1+\Delta_3-\Delta_2)<d, \\ \nonumber
{\rm Re}(\bar{\Delta}_1+\Delta_5-\Delta_6)<d, \qquad {\rm Re}(\bar{\Delta}_1+\Delta_6-\Delta_5)<d.
\ee
Note that these conditions are satisfied for the type I representations.

We see that only 4 of the quantities in the denominator involve $x_1$. In order to take the integral, let us use the
Feynman representation for them. In other words, let us consider:
\be\label{func-I}
I_{\Delta_1,\Delta_2,\Delta_3,\Delta_5,\Delta_6}(x_2,x_3,x_5,x_6) =
\int_0^\infty dt_{12} dt_{13} dt_{15} dt_{16} \\ \nonumber
t_{12}^{\frac{\Delta_1+\Delta_2-\Delta_3}{2}-1} t_{13}^{\frac{\Delta_1+\Delta_3-\Delta_2}{2}-1}
t_{15}^{\frac{\bar{\Delta}_1+\Delta_5-\Delta_6}{2}-1} t_{16}^{\frac{\bar{\Delta}_1+\Delta_6-\Delta_5}{2}-1} \\ \nonumber
\int_{S^d} d^dx_1 e^{-t_{12} x_{12}^2 -t_{13} x_{13}^2  - t_{15} x_{15}^2 -t_{16} x_{16}^2},
\ee
In terms of this function:
\be\label{D-I}
D_{\Delta_1,\Delta_2,\Delta_3,\Delta_5,\Delta_6}(x_2,x_3,x_5,x_6) = \frac{I_{\Delta_1,\Delta_2,\Delta_3,\Delta_5,\Delta_6}(x_2,x_3,x_5,x_6)}
{(x_{23})^{\Delta_2+\Delta_3-\Delta_1}(x_{56})^{\Delta_5+\Delta_6-\bar{\Delta}_1}}
\\ \nonumber
\frac{C_{\Delta_1,\Delta_2,\Delta_3}C_{\bar{\Delta}_1,\Delta_5,\Delta_6}}
{\Gamma(\frac{\Delta_1+\Delta_2-\Delta_3}{2})\Gamma(\frac{\Delta_1+\Delta_3-\Delta_2}{2})
\Gamma(\frac{\bar{\Delta}_1+\Delta_5-\Delta_6}{2})\Gamma(\frac{\bar{\Delta}_1+\Delta_6-\Delta_5}{2})}.
\ee

Let us now evaluate the function \eqref{func-I}. The integral over $x_1$ is taken using \eqref{2}:
\be
\int_{S^d} d^dx_1 e^{-t_{12} x_{12}^2 -t_{13} x_{13}^2  - t_{15} x_{15}^2 -t_{16} x_{16}^2} = \\ \nonumber
\frac{\pi^{d/2}}{(S_t^1)^{d/2}}  e^{-(t_{12} t_{13} x_{23}^2 + t_{12} t_{15} x_{25}^2 + t_{12} t_{16} x_{26}^2
+ t_{13} t_{15} x_{35}^2 + t_{13} t_{16} x_{36}^2 + t_{15} t_{16} x_{56}^2)/S_t^1},
\ee
where
\be
S_t^1 = t_{12}+t_{13}+t_{15}+t_{16}.
\ee

Let us now make a series of changes of variables. First, let us choose:
\be
t_{1i}=(S_t^1)^{1/2} t_{1i}', \qquad {\rm det} \left( \frac{\partial t_{1i}}{\partial t_{1i}'} \right) = 2 (S_t^1)^2.
\ee
After this change of variables we get:
\be
 I_{\Delta_1,\Delta_2,\Delta_3,\Delta_5,\Delta_6}(x_2,x_3,x_5,x_6)  =2\pi^{d/2} \int_0^\infty dt_{12} dt_{13} dt_{15} dt_{16} \\ \nonumber
t_{12}^{\frac{\Delta_1+\Delta_2-\Delta_3}{2}-1} t_{13}^{\frac{\Delta_1+\Delta_3-\Delta_2}{2}-1}
t_{15}^{\frac{\bar{\Delta}_1+\Delta_5-\Delta_6}{2}-1} t_{16}^{\frac{\bar{\Delta}_1+\Delta_6-\Delta_5}{2}-1} \\ \nonumber
e^{-t_{12} t_{13} x_{23}^2 - t_{12} t_{15} x_{25}^2 - t_{12} t_{16} x_{26}^2
- t_{13} t_{15} x_{35}^2 - t_{13} t_{16} x_{36}^2 - t_{15} t_{16} x_{56}^2}.
\ee

The second change of variables is:
\be
t_{12} t_{13} \to \frac{t_{12} t_{13}}{x_{23}^2}, \, t_{12} t_{15} \to \frac{t_{12} t_{15}}{x_{25}^2},
\, t_{12} t_{16} \to \frac{t_{12} t_{16}}{x_{26}^2}, \, t_{13} t_{15} \to \frac{t_{13} t_{15}}{x_{35}^2}.
\ee
As the consequence we get:
\be
t_{13} t_{16} = \frac{t_{13} t_{15} t_{12} t_{16}}{t_{12} t_{15}} \to \frac{x_{25}^2}{x_{35}^2 x_{26}^2} t_{13} t_{16}, \, 
t_{15} t_{16} = \frac{t_{13} t_{15} t_{12} t_{16}}{t_{12} t_{13}} \to \frac{x_{23}^2}{x_{35}^2 x_{26}^2} t_{15} t_{16}.
\ee
Or, equivalently:
\be
t_{12}\to t_{12} \frac{x_{35}}{x_{23} x_{25}}, \,\, t_{13}\to t_{13} \frac{x_{25}}{x_{23} x_{35}}, \,\,
t_{15}\to t_{15} \frac{x_{23}}{x_{25} x_{35}}, \,\, t_{16}\to t_{16} \frac{x_{25} x_{23}}{x_{26}^2 x_{35}}.
\ee
As the result of this transformation we get:
\be\label{I-1}
 I_{\Delta_1,\Delta_2,\Delta_3,\Delta_5,\Delta_6}(x_2,x_3,x_5,x_6) = \qquad\qquad \\ \nonumber 
\frac{2\pi^{d/2}}{(x_{23})^{\Delta_1-\bar{\Delta}_1} (x_{25})^{\Delta_2-\Delta_3+\Delta_5-\Delta_6} 
(x_{26})^{\bar{\Delta}_1+\Delta_6-\Delta_5} (x_{35})^{\bar{\Delta}_1+\Delta_3-\Delta_2}} \\ \nonumber 
\int_0^\infty dt_{12} dt_{13} dt_{15} dt_{16} \,\, e^{-t_{12} t_{13} - t_{12} t_{15} - t_{12} t_{16} 
- t_{13} t_{15}  - t_{13} t_{16} u - t_{15} t_{16} v} \\ \nonumber
t_{12}^{\frac{\Delta_1+\Delta_2-\Delta_3}{2}-1} t_{13}^{\frac{\Delta_1+\Delta_3-\Delta_2}{2}-1}
t_{15}^{\frac{\bar{\Delta}_1+\Delta_5-\Delta_6}{2}-1} t_{16}^{\frac{\bar{\Delta}_1+\Delta_6-\Delta_5}{2}-1}.
\ee
Here we have introduced the cross-ratios:
\be\label{cross}
u=\frac{x_{25}^2 x_{36}^2}{x_{26}^2 x_{35}^2}, \qquad v=\frac{x_{23}^2 x_{56}^2}{x_{26}^2 x_{35}^2}.
\ee
The two cross-ratios are related for $d=2$ but independent for $d>2$. 

We now specialize to the type I representations. 

To compute the integrals over the Feynman parameters we, following an analogous computation in \cite{Gleb}, 
use the Mellin-Barnes integral representation:
\be\label{barnes}
e^{-z} = \frac{1}{2\pi i} \int_{r-i\infty}^{r+i\infty} ds \, 
\Gamma(-s) z^s, \qquad r<0, \qquad |{\rm arg} z|<\frac{\pi}{2}.
\ee
Thus, the integral over Feynman parameters becomes:
\be\label{eq:interchange}
\frac{1}{2} I_{\Delta_1,\Delta_2,\Delta_3,\Delta_5,\Delta_6}(u,v):=  
\int \frac{d\lambda ds}{(2\pi i)^2} \Gamma(-\lambda) \Gamma(-s) u^{s}  v^{\lambda}\\ \nonumber
\int_0^\infty dt_{12} dt_{13} dt_{15} dt_{16}\,\, e^{-t_{12} t_{13} - t_{12} t_{15} - t_{12} t_{16} 
- t_{13} t_{15}} \\ \nonumber
t_{12}^{\frac{\Delta_1+\Delta_2-\Delta_3}{2}-1} t_{13}^{\frac{\Delta_1+\Delta_3-\Delta_2}{2}+s-1}
t_{15}^{\frac{\bar{\Delta}_1+\Delta_5-\Delta_6}{2}+\lambda-1} t_{16}^{\frac{\bar{\Delta}_1+\Delta_6-\Delta_5}{2}+s+\lambda-1}. 
\ee
To justify the interchange of integrals that has led to \eqref{eq:interchange}, 
we assume that the parameter $r$ in the two 
Mellin-Barnes contours is the same and satisfies $-d/8 < r < 0$. 
This guarantees that the exponent of each $t_{ij}$ in \eqref{eq:interchange} has real part greater than~$-1$. 

In the Feynman parameter integrals, we change variables by: 
\be
t_{12} t_{13} = u_1, \,\, t_{12} t_{15} =u_2, \,\, t_{12} t_{16}=u_3, \,\, t_{13} t_{15}=u_4,
\ee
or, equivalently:
\be
t_{12}^2=\frac{u_1 u_2}{u_4}, \,\, t_{13}^2=\frac{u_1 u_4}{u_2}, \,\, t_{15}^2 = \frac{u_2 u_4}{u_1}, \,\, t_{16}^2 = u_3^2 \frac{u_4}{u_1 u_2}.
\ee
The Jacobian of this transformation is:
\be 
{\rm det}\left( \frac{\partial t_{1i}}{\partial u_j} \right) = \frac{1}{2u_1 u_2}.
\ee
Therefore, we get:
\be\label{I-2}
 I_{\Delta_1,\Delta_2,\Delta_3,\Delta_5,\Delta_6}(u,v)= \int \frac{d\lambda ds}{(2\pi i)^2} \Gamma(-\lambda) \Gamma(-s) u^{s}  v^{\lambda}
 \\ \nonumber 
\Gamma(\alpha+s+\lambda)\Gamma(\beta+s+\lambda) \Gamma(\gamma-s) \Gamma(\delta-\lambda),
\ee
where we have introduced the following quantities:
\be\nonumber
\alpha=\frac{\bar{\Delta}_1+\Delta_3-\Delta_2}{2}, \\ \label{alpha-b}
\beta=\frac{\bar{\Delta}_1+\Delta_6-\Delta_5}{2}, \\ \nonumber
\gamma=\frac{\Delta_2-\Delta_3+\Delta_5-\Delta_6}{2}, \\ \nonumber
\delta=\frac{\Delta_1-\bar{\Delta}_1}{2}.
\ee

Combining this, and the expressions \eqref{I-1}, \eqref{D-I}, we have for the 4-point function:
\be\label{D-fin}
D_{\Delta_1,\Delta_2,\Delta_3,\Delta_5,\Delta_6}(x_2,x_3,x_5,x_6) = K_{\Delta_1,\Delta_2,\Delta_3,\Delta_5,\Delta_6} 
\frac{ I_{\Delta_1,\Delta_2,\Delta_3,\Delta_5,\Delta_6}(u,v)}{(x_{25})^{\Delta_2-\Delta_3+\Delta_5-\Delta_6}}
\\ \nonumber
\frac{1}{(x_{23})^{\Delta_2+\Delta_3-\bar{\Delta}_1} 
(x_{26})^{\bar{\Delta}_1+\Delta_6-\Delta_5} (x_{35})^{\bar{\Delta}_1+\Delta_3-\Delta_2}(x_{56})^{\Delta_5+\Delta_6-\bar{\Delta}_1}},
\ee
where we have introduced:
\be\label{K}
K_{\Delta_1,\Delta_2,\Delta_3,\Delta_5,\Delta_6} = 
\frac{\pi^{d/2} C_{\Delta_1,\Delta_2,\Delta_3}C_{\bar{\Delta}_1,\Delta_5,\Delta_6}}
{\Gamma(\frac{\Delta_1+\Delta_2-\Delta_3}{2})\Gamma(\frac{\Delta_1+\Delta_3-\Delta_2}{2})
\Gamma(\frac{\bar{\Delta}_1+\Delta_5-\Delta_6}{2})\Gamma(\frac{\bar{\Delta}_1+\Delta_6-\Delta_5}{2})}.
\ee
Thus, essentially all non-triviality of the 4-point function is in the dependence of the function
$I_{\Delta_1,\Delta_2,\Delta_3,\Delta_5,\Delta_6}(u,v)$ given by \eqref{I-2} on the two
cross-ratios $u,v$. The integral representation \eqref{I-2} that we have obtained
may be used to derive 
power series expansions of $I_{\Delta_1,\Delta_2,\Delta_3,\Delta_5,\Delta_6}(u,v)$ 
in $u$ and $v$ \cite{Gleb}.

\section{Computation of the Racah-Wigner coefficients}
\label{sec:comp}

Having obtained the 4-point function we can use this result to compute the RW coefficients. 
Recalling \eqref{6j-4-point},
we get:
\be\label{6j-interm}
(6\Delta)=K_{\Delta_1,\Delta_2,\Delta_3,\Delta_5,\Delta_6} K_{\Delta_4,\bar{\Delta}_2,\bar{\Delta}_3,\bar{\Delta}_5,\bar{\Delta}_6}
\int_{S^d} \frac{d^dx_2 d^dx_3 d^dx_5 d^dx_6}{x_{23}^d x_{26}^d x_{35}^d x_{56}^d} \,\, \\ \nonumber 
\left( \frac{1}{v}\right)^{\frac{\Delta_1-\Delta_4+\Delta_2+\Delta_5-d}{2}} 
I_{\Delta_1,\Delta_2,\Delta_3,\Delta_5,\Delta_6}(u,v) I_{\Delta_4,\bar{\Delta}_2,\bar{\Delta}_6,\bar{\Delta}_5,\bar{\Delta}_3}(u/v,1/v).
\ee
Here we have used the fact that under the exchange of $x_3, x_6$ we
have: $v\to1/v, u\to u/v$. We have also used the relation \eqref{sum}
between the conformal dimension and its dual and combined all powers
of $x_{ij}$. Note that the integration measure 
in \eqref{6j-interm} is exactly right to 
give an invariant expression, as this measure can be verified to be invariant
under conformal transformations of the $x_i$.

What remains is to compute the sphere integrals in \eqref{6j-interm}. This integral is:
\be\label{int-2}
I_d(a,b)=\int_{S^d} \frac{dx_2 dx_3 dx_5 dx_6}{x_{23}^d x_{56}^d x_{26}^d x_{35}^d}
\left(\frac{x_{23}x_{56}}{x_{26} x_{35}}\right)^{2a} \left(\frac{x_{25}x_{36}}{x_{26} x_{35}}\right)^{2b},
\ee
where 
\be\label{a-b}
a=- A+\lambda-s'-\lambda', \qquad b=s+s', \qquad A= \frac{\Delta_1-\Delta_4+\Delta_2+\Delta_5-d}{2}
\ee
As it stands, the integral \eqref{int-2} 
still diverges because of the volume of the gauge group. 
We eliminate this divergence by the replacement \eqref{measure-d}, 
leaving $x_2$ to be integrated over but 
fixing the locations of the other three points. 
The integral becomes:
\be\label{I-d-2}
I_d(a,b)= |x^0_3-x^0_6|^d \left|\frac{x^0_5-x^0_6}{x^0_3-x^0_5}\right|^{2a} 
\left|\frac{x^0_3-x^0_6}{x^0_3-x^0_5}\right|^{2b} \tilde{I}_d(a,b),
\ee
where
\be\label{eq:I-gend}
\tilde{I}_d(a,b)=\int_{S^d} 
dx_2\,\frac{|x_2-x^0_3|^{2a-d} |x_2-x^0_5|^{2b}}{|x_2-x^0_6|^{2a+2b+d}}.
\ee
The convergence conditions in \eqref{eq:I-gend} are 
\be
{\rm Re}(a)>0, \qquad {\rm Re}(b+d/2)>0, \qquad {\rm Re}(a+b)<0, 
\ee
and they are satisfied for our Mellin-Barnes contour choice, $-d/8 < r < 0$. 

To compute the integral over $x_2$ we use the Feynman parametrization. We have:
\be\label{I-d-1}
\tilde{I}_d(a,b)= \frac{1}{\Gamma(d/2-a)\Gamma(-b)\Gamma(a+b+d/2)} \int_{S^d} 
dx_2\, t_3^{d/2-a-1} t_5^{-b-1} t_6^{a+b+d/2-1} \\ \nonumber e^{-t_3|x_2-x^0_3|^2 -t_5|x_2-x^0_5|^2-t_6|x_2-x^0_6|^2}.
\ee
We have already dealt with essentially the same integral in section \ref{sec:3j}, formula \eqref{3j-1}.
Thus, we shall be sketchy here. One first takes the $x_2$ integral using \eqref{2}. One then makes a
series of rescalings of Feynman parameters $t_i$. The powers of these parameters in 
\eqref{I-d-1} are such that the rescaling \eqref{change-1} completely removes the quantity
$S_t$ from the integral. Rescaling \eqref{change-2} takes the differences $|x^0_i-x^0_j|$ out
of the integral, and their resulting powers are exactly such that they cancel the similar
quantities in \eqref{I-d-2}. The last change of variables \eqref{change-3} reduces all the
integrals to those giving $\Gamma$-functions. The final result is:
\be\label{I-d-3}
I_d(a,b)=\frac{\pi^{d/2} \Gamma(a)\Gamma(d/2+b)\Gamma(-a-b)}{\Gamma(d/2-a)\Gamma(-b)\Gamma(a+b+d/2)}=
\frac{\pi^{d/2} \Upsilon_d(a) \Upsilon_d(-a-b)}{\Upsilon_d(-b)},
\ee
where we have introduced a dimension-dependent $\Upsilon$-function given by:
\be
\Upsilon_d(x)=\frac{\Gamma(x)}{\Gamma(d/2-x)}.
\ee
Substituting the values \eqref{a-b} of the parameters $a$ and~$b$, 
our final result for the $(6\Delta)$ symbol is: 
\be\label{6j-d}
(6\Delta)=
K_{\Delta_1,\Delta_2,\Delta_3,\Delta_5,\Delta_6} K_{\Delta_4,\bar{\Delta}_2,\bar{\Delta}_3,\bar{\Delta}_5,\bar{\Delta}_6}
\\ \nonumber
\pi^{d/2} \int \frac{d\lambda ds}{(2\pi i)^2} \int \frac{d\lambda' ds'}{(2\pi i)^2} \Gamma(-s) \Gamma(-\lambda) \Gamma(-s') \Gamma(-\lambda') \\ \nonumber
 \Gamma(\gamma-s) \Gamma(\delta-\lambda) \Gamma(\alpha+s+\lambda)\Gamma(\beta+s+\lambda)\\ \nonumber
\Gamma(\gamma'-s') \Gamma(\delta'-\lambda') \Gamma(\alpha'+s'+\lambda')\Gamma(\beta'+s'+\lambda')\\ \nonumber
\frac{\Upsilon_d(\lambda-s'-\lambda'-A)\Upsilon_d(\lambda'-s-\lambda+A)}{\Upsilon_d(-s-s')}.
\ee
Using Stirling's formula and the assumption $-d/8<r<0$ for the
Mellin-Barnes contours, it can be verified that the quadruple integral
in \eqref{6j-d} is convergent in absolute value. Equation \eqref{6j-d}
therefore gives a manifestly well-defined expression for the
$(6\Delta)$ symbol.

\section*{Acknowledgments}

We thank John Barrett and Eli Hawkins for insightful discussions
and an anonymous referee for pointing out the presence of a 
multiplicity label for $d=1$. 
KK is supported by an EPSRC Advanced Fellowship. 

\appendix

\section{Integrals}
\label{app:integrals}

In this appendix we give the formulas that are central to the methods of integration that we use. 
The same method was exploited
in the context of AdS/CFT correspondence, see e.g. \cite{Gleb}. 
The standard Feynman parameter method is based on the following representation:
\be\label{feyn}
\frac{1}{z^\lambda}=\frac{1}{\Gamma(\lambda)} \int_0^\infty dt t^{\lambda-1} e^{-t z}.
\ee

We shall also need the following two integrals:
\be\label{1}
\int_0^\infty \frac{d\xi_0}{\xi_0^{d+1}}\, \xi_0^{\sum_i \Delta_i} e^{-\sum_i t_i \xi_0^2} = 
\frac{1}{2} \left( S_t \right)^{\frac{d-\sum_i \Delta_i}{2}} \Gamma(\frac{\sum_i \Delta_i-d}{2}),
\ee
and 
\be\label{2}
\int_{S^d} d^dx e^{-\sum_i t_i |x-x_i|^2} = \frac{\pi^{d/2}}{S_t^{d/2}} 
e^{-\frac{1}{S_t}(\sum_{i<j} t_i t_j |x_i-x_j|^2)}.
\ee
In both of these formulas:
\be
S_t = \sum_i t_i.
\ee

\section{Gauge-fixed integration measure}
\label{app:measure}

In this appendix we verify that the gauge-fixing procedure 
\eqref{measure-d} results from decomposing the Haar measure on 
$\SO_0(1,d+1)$ as stated in section~\ref{sec:norm}. 
We assume $d>1$ as in the main text.

\subsection{Notation}

${}$ 

We first recall some basic properties 
of the conformal group~\cite{Dobrev}. 

We view $\SO_0(1,d+1)$ as the matrix group in its defining
representation, acting on $\RR^{1,d+1}$ by matrix multiplication of a
column vector. We write the points in 
$\RR^{1,d+1}$ 
as ${(t,y,z^1, \ldots , z^d)} =: {(t,y,z)}$,
where $z$ is in $\RR^d$. 

The action
of $\SO_0(1,d+1)$ as the conformal group on $\RR^d \cup \{\infty\}$ is
obtained from that on the future null cone in $\RR^{1,d+1}$, $t =
\sqrt{y^2 + z^2}$, by parametrising the cone as $(t,y,z) = s\bigl
(\frac12(1+x^2), \frac12(1-x^2), x \bigr)$, where $s\ge0$ and $x\in
\RR^d \cup \{\infty\}$. The point $x=\infty$ corresponds to the null
ray $t+y=0=z$.

If $q\in\RR^d$ 
and $\lambda>0$, we 
define the $\SO_0(1,d+1)$ matrices 
\be
N_+(q) := \left(
\begin{array}{ccc}
1 + \frac12 q^2 & \frac12 q^2 & q^T 
\\
-\frac12 q^2 & 1 - \frac12 q^2 & - q^T 
\\
q & q & \II_d
\end{array}
\right) , 
\ee
\be
N_-(q) := \left(
\begin{array}{ccc}
1 + \frac12 q^2 & - \frac12 q^2 & q^T 
\\
\frac12 q^2 & 1 - \frac12 q^2 & q^T 
\\
q & -q & \II_d
\end{array}
\right) , 
\ee
\be
A(\lambda) := \left(
\begin{array}{ccc}
\frac12 (\lambda^{-1} + \lambda) & \frac12  (\lambda^{-1} - \lambda) & 0 
\\
\frac12  (\lambda^{-1} - \lambda) & \frac12 (\lambda^{-1} + \lambda) & 0 
\\
0 & 0 & \II_d
\end{array}
\right), 
\ee
where $q$ is viewed as a column vector and 
$\II_d$ is the $d \times d$ unit matrix. 
In terms of the action on $\RR^{1,d+1}$, 
$N_{\pm}(q)$ are null rotations and $A(\lambda)$ is a boost. 
The conformal action of 
$N_+(q)$ on $\RR^d$ is the translation $x \mapsto x + q$, 
the action of $N_-(q)$ is the proper conformal transformation 
$x/(x^2) \mapsto x/(x^2) + q$, 
and the action of $A(\lambda)$ is the dilatation $x \mapsto \lambda x$. 

Finally, if $u$ and $v$ are two unit vectors in $\RR^d$, $u\ne-v$, 
we define the $\SO(d)$ matrix 
\be
\label{eq:tildeS}
\tilde{S}(v,u):= 
\II_d + 2 \, v \otimes u^T
- \frac{( u + v ) \otimes (u + v )^T}{1 + u \cdot v}  . 
\ee
$\tilde{S}(v,u)$ is a rotation in the plane of 
$u$ and $v$ and takes $u$ to~$v$, 
reducing to $\II_d$ when $u=v$. 
Given $\tilde{S}(v,u)$, we define the corresponding block-diagonal 
$\SO_0(1,d+1)$ matrix by $S(v,u) := 
\text{diag} \bigl(\II_2, \tilde{S}(v,u) \bigr)$.

\subsection{Parametrisation}

${}$

Let $x_A$, $x_B$ and $x_C$ be three distinct points in 
$\RR^d$. We set 
\be
x_D := 
\frac{x_{AB}^2 x_{AC}^2}{x_{BC}^2}
\left(
\frac{x_B - x_A}{x_{AB}^2}
- 
\frac{x_C - x_A}{x_{AC}^2}
\right) , 
\ee
where $x_{ij} := |x_i - x_j|$ as in the main text. As $|x_D|
= x_{AB} x_{AC} x_{BC}^{-1}$, it follows that $x_D \ne 0$. 

Let 
$v$ be a unit vector in $\RR^d$. 
When $x_D/|x_D| \ne -v$, we define 
\be
\label{eq:h-def}
h(x_A,x_B,x_C) 
:= 
S(v,x_D/|x_D|)
A (1/|x_D|) 
N_-(b) N_+(a) , 
\ee
where $a := -x_A$ and $b := x_{AC}^{-2}(x_A - x_C)$. 
A direct computation shows that the conformal action of $h$ takes 
the triple $(x_A,x_B,x_C)$ to the triple $(0,v,\infty)$. 
Each $\SO_0(1,d+1)$ matrix $g$ whose conformal action takes 
$(x_A,x_B,x_C)$ to $(0,v,\infty)$ 
can therefore be uniquely
written as 
\be
\label{eq:g-decomp} 
g = R h , 
\ee
where 
$R= \text{diag} \bigl(\II_2, \tilde{R} \bigr)$ and 
$\tilde{R}$ is an $\SO(d)$ matrix in the 
$\SO(d-1)$ subgroup that fixes~$v$. 
Note that this subgroup is trivial for $d=2$ but nontrivial for
$d>2$. 

If $v$ is considered fixed, the above discussion shows that formulas
\eqref{eq:h-def}
and \eqref{eq:g-decomp} give a unique parametrisation of an open subset
in $\SO_0(1,d+1)$ by the triple $(x_A,x_B,x_C)$ 
and a group isomorphic to $\SO(d-1)$. 
This parametrisation does not cover all of $\SO_0(1,d+1)$, but the
missing subset is of measure zero and can be recovered 
by choosing a different $v$ and by limits
in which one of $x_A$, $x_B$ and $x_C$ is taken to infinity. 

Now, the right action of $\SO_0(1,d+1)$ on
\eqref{eq:g-decomp} induces on the triple $(x_A,x_B,x_C)$
the conformal $\SO_0(1,d+1)$ action of the main text. To eliminate
the volume divergence of the main text, we therefore need to write the
Haar measure on $\SO_0(1,d+1)$ in the
decomposition \eqref{eq:g-decomp} and identify the part of the measure
that is associated with the
noncompact
factor~$h$. If this measure is 
$\mu(x_A,x_B,x_C) \, d^dx_A \, d^dx_B \, 
d^dx_C$, the divergence will
be
eliminated by the replacement 
\be
d^d x_A \, d^d x_B \, d^d x_C \, 
\to 
\frac{\delta^d(x_A-x_A^0)\delta^d(x_B-x_B^0)\delta^d(x_C-x_C^0)}
{\mu(x_A,x_B,x_C)}
\, d^d x_A \, d^d x_B \, d^d x_C , 
\ee
where the distinct points $x_A^0$, $x_B^0$ and $x_C^0$ can be chosen
arbitrarily. 

\subsection{Measure computation}

${}$

From 
\eqref{eq:h-def}
and 
\eqref{eq:g-decomp} we find 
\be
\label{eq:dg-ginv}
{(AR)}^{-1} \bigl( dg \, g^{-1} \bigr) (AR) 
=  R^{-1}  dR  + dS \, S^{-1} + S Q S^{-1},
\ee
where 
\be
\label{eq:Q-def}
Q &=& 
\left( {|x_D|}^{-1} d{(|x_D|)} + 2 b\cdot da \right) 
\left(
\begin{array}{ccc}
0
& 
1
&
0
\\
1
&
0 
&
0
\\
0
&
0
&
0
\end{array}
\right)
\\
&&
\nonumber
+ 
\left(
\begin{array}{ccc}
0
& 
0
&
(1+b^2) da^T - 2(b\cdot da) b^T + db^T
\\
0
&
0 
&
0
\\
(1+b^2) da - 2(b\cdot da) b + db
&
0
&
0
\end{array}
\right)
\\
\nonumber
&&
+ 
\left(
\begin{array}{ccc}
0
& 
0
&
0
\\
0
&
0 
&
(1-b^2) da^T + 2(b\cdot da) b^T - db^T
\\
0
&
(1-b^2) da + 2(b\cdot da) b - db
&
0
\end{array}
\right)
\\
\nonumber
&&
+ 
\left(
\begin{array}{ccc}
0
& 
0
&
0
\\
0
&
0 
&
0
\\
0
&
0
&
2 \bigl( b \otimes da^T - da \otimes b^T \bigr)
\end{array}
\right) . 
\ee
The explicit form of the term $dS \, S^{-1}$ can be found
from~\eqref{eq:tildeS}.

To find the part of the Haar measure that corresponds to the
noncompact factor~$h$, we project the right-hand side of
(\ref{eq:dg-ginv}) to the subspace orthogonal (with respect to the
Killing form) to the subspace generated by $R^{-1} dR$. Choosing
$v$ for concreteness to be $(1,0,0,\ldots)$, this amounts to setting
the lower-right $(d-1)\times(d-1)$ block to zero but leaving the other
components unchanged. From the coefficients of the remaining $3d$ Lie
algebra elements we can then identify the $3d \times 3d$ determinant
that gives the desired measure. Evaluating the determinant by
elementary matrix algebra techniques, we find that this measure is
\be
\label{measure}
\frac{d^d x_A \, d^d x_B \, d^d x_C}{{|x_A-x_B|}^d 
{|x_B-x_C|}^d {|x_C-x_A|}^d} . 
\ee
Hence the gauge fixing procedure 
\eqref{measure-d} follows. 

As a check, it can be directly verified that the measure
\eqref{measure} is invariant under the conformal action of
$\SO_0(1,d+1)$.

\end{document}